\def\pz{\pi^0}
\def\g{\gamma}
\def\ee{e^+e^-}
\def\etal{{\it et. al}}

\documentstyle[prl,aps]{revtex}

\input{psfig}

\begin{document}
%
%
\title{Measurement of the decay $K_L\to\pz\g\g$}
\author{\parindent=0.in
The KTeV Collaboration\\
A.~Alavi-Harati$^{12}$,
I.F.~Albuquerque$^{10}$,
T.~Alexopoulos$^{12}$,
M.~Arenton$^{11}$,
K.~Arisaka$^2$,
S.~Averitte$^{10}$,
A.R.~Barker$^5$,
L.~Bellantoni$^7$,
A.~Bellavance$^9$,
J.~Belz$^{10}$,
R.~Ben-David$^7$,
D.R.~Bergman$^{10}$,
E.~Blucher$^4$, 
G.J.~Bock$^7$,
C.~Bown$^4$, 
S.~Bright$^4$,
E.~Cheu$^{1,\dagger}$,
S.~Childress$^7$,
R.~Coleman$^7$,
M.D.~Corcoran$^9$,
G.~Corti$^{11}$, 
B.~Cox$^{11}$,
M.B.~Crisler$^7$,
A.R.~Erwin$^{12}$,
R.~Ford$^7$,
A.~Golossanov$^{11}$,
G.~Graham$^4$, 
J.~Graham$^4$,
K.~Hagan$^{11}$,
E.~Halkiadakis$^{10}$,
K.~Hanagaki$^8$,  
S.~Hidaka$^8$,
Y.B.~Hsiung$^7$,
V.~Jejer$^{11}$,
J.~Jennings$^2$,
D.A.~Jensen$^7$,
R.~Kessler$^4$,
H.G.E.~Kobrak$^{3}$,
J.~LaDue$^5$,
A.~Lath$^{10}$,
A.~Ledovskoy$^{11}$,
P.L.~McBride$^7$,
A.P.~McManus$^{11}$,
P.~Mikelsons$^5$,
E.~Monnier$^{4,*}$,
T.~Nakaya$^7$,
U.~Nauenberg$^5$,
K.S.~Nelson$^{11}$,
H.~Nguyen$^7$,
V.~O'Dell$^7$, 
M.~Pang$^7$, 
R.~Pordes$^7$,
V.~Prasad$^4$, 
C.~Qiao$^4$, 
B.~Quinn$^4$,
E.J.~Ramberg$^7$, 
R.E.~Ray$^7$,
A.~Roodman$^4$, 
M.~Sadamoto$^8$, 
S.~Schnetzer$^{10}$,
K.~Senyo$^8$, 
P.~Shanahan$^7$,
P.S.~Shawhan$^4$, 
W.~Slater$^2$,
N.~Solomey$^4$,
S.V.~Somalwar$^{10}$, 
R.L.~Stone$^{10}$, 
I.~Suzuki$^8$,
E.C.~Swallow$^{4,6}$,
R.A.~Swanson$^{3}$,
S.A.~Taegar$^1$,
R.J.~Tesarek$^{10}$, 
G.B.~Thomson$^{10}$,
P.A.~Toale$^5$,
A.~Tripathi$^2$,
R.~Tschirhart$^7$, 
Y.W.~Wah$^4$,
J.~Wang$^1$,
H.B.~White$^7$, 
J.~Whitmore$^7$,
B.~Winstein$^4$, 
R.~Winston$^4$, 
J.-Y.~Wu$^5$,
T.~Yamanaka$^8$,
E.D.~Zimmerman$^4$\\
\vspace*{0.1in}
\footnotesize
$^1$ University of Arizona, Tucson, Arizona 85721 \\
$^2$ University of California at Los Angeles, Los Angeles, California 90095 \\
$^{3}$ University of California at San Diego, La Jolla, California 92093 \\
$^4$ The Enrico Fermi Institute, The University of Chicago, 
Chicago, Illinois 60637 \\
$^5$ University of Colorado, Boulder, Colorado 80309 \\
$^6$ Elmhurst College, Elmhurst, Illinois 60126 \\
$^7$ Fermi National Accelerator Laboratory, Batavia, Illinois 60510 \\
$^8$ Osaka University, Toyonaka, Osaka 560 Japan \\
$^9$ Rice University, Houston, Texas 77005 \\
$^{10}$ Rutgers University, Piscataway, New Jersey 08855 \\
$^{11}$ The Department of Physics and the Institute of Nuclear and 
Particle Physics, University of Virginia, 
Charlottesville, Virginia 22901 \\
$^{12}$ University of Wisconsin, Madison, Wisconsin 53706 \\
$^{*}$ On leave from C.P.P. Marseille/C.N.R.S., France \\
$^\dagger$ To whom correspondence should be addressed. Electronic
address: elliott@physics.arizona.edu\\
\normalsize
}

\maketitle
\begin{abstract}

We report on a new measurement of the decay
$K_L\to\pz\g\g$ by the KTeV experiment at Fermilab. 
We determine the $K_L\to\pz\g\g$ branching ratio to be
$(1.68 \pm 0.07\pm 0.08)\times 10^{-6}$. Our data shows the first
evidence for a low-mass $\g\g$ signal as predicted by
recent ${\cal O}(p^6)$ chiral perturbation calculations that include 
vector meson exchange contributions. From our data, we extract a 
value for the effective vector coupling 
$a_V = -0.72\pm 0.05 \pm 0.06$.\\
\vspace*{0.1in}
\end{abstract}

\newpage
\narrowtext
\twocolumn

\parindent=0.3in
\normalsize

Studying the decay $K_L\to\pz\g\g$ is important 
for understanding the low energy hadron dynamics of
chiral perturbation theory. Because of the high backgrounds
associated with this decay, the first measurement of 
the $K_L\to\pz\g\g$ branching ratio\cite{gbarr1} was performed only relatively
recently. Other measurements quickly confirmed this 
result\cite{vaia,gbarr_pi0gg} with the measured branching ratio
approximately three times higher than the predictions from 
${\cal O}(p^4)$ chiral perturbation calculations\cite{EPR}.
More recent calculations\cite{ecker,donoghue,dambrosio,sehgal}
which include ${\cal O}(p^6)$ corrections and vector meson exchange
contributions obtain a branching ratio consistent with the measured value. 

\begin{figure}[hbt]
  \centerline{
    \psfig{file=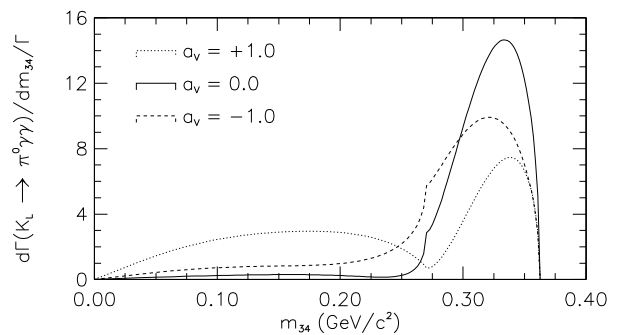,width=8.5cm}}
  \caption{Theoretical predictions for $m_{34}$ for various
           values of $a_V$.}
  \label{fig:av}
\end{figure}

The rate for $K_L\to\pz\g\g$ can be expressed in terms of
two independent Lorentz invariant amplitudes
which represent $J=0$ and $J=2$ two-photon states. The 
contributions of the two amplitudes are determined by two Dalitz parameters,
$z = ({m_{34}/ m_K})^2$ and $y = | E_3 - E_4 |/m_K$, where
$E_3$ and $E_4$ are the energies of the non-$\pz$ photons 
in the kaon center of mass frame and $m_{34}$ is their invariant
mass. Vector meson exchange contributions
to the amplitudes are parametrized by an effective coupling constant
$a_V$\cite{sehgal} and 
the $z$ and $y$ Dalitz variables are sensitive to the value of $a_V$
as shown in Figure \ref{fig:av}.
In particular, certain values of $a_V$ result in a sizeable low-mass
tail in the $m_{34}$ distribution. Due to limited statistics,
previous measurements were not sensitive to such a low-mass
tail. With a sufficiently large event sample, 
it is possible to test the predictions
of ${\cal O}(p^6)$ chiral perturbation theory with
vector meson contributions and to precisely determine the parameter
$a_V$ directly from the data. A precise determination of $a_V$ also
allows one to predict the relative contributions of the
CP conserving and direct CP violating components of $K_L\to\pz\ee$
which can proceed through the CP conserving process
$K_L\to\pz\g^*\g^*\to\pz\ee$.

\begin{figure}[hbt]
  \centerline{
    \psfig{file=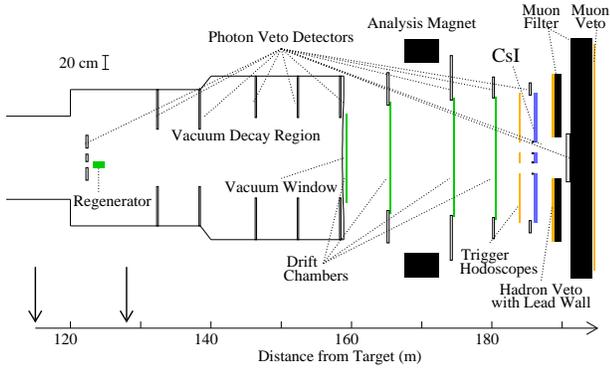,width=8.5cm}}
  \caption{Plan view of the KTeV detector as configured for the E832
           experiment. The arrows indicate the decay region considered
           for this analysis.}
  \label{fig:det}
\end{figure}

%
%
We recorded $K_L\to\pz\g\g$ events using the KTeV detector
located at Fermilab. 
Figure~\ref{fig:det} shows a plan view of the
detector as it was configured for the E832 experiment. The main goal
of the E832 experiment is to search for direct CP violation in
$K\to\pi\pi$ decays. 
The $K_L\to\pz\g\g$ analysis utilizes the data
set collected by the KTeV experiment during 1996 and 1997 for the
direct CP violation search.
Neutral kaons are produced in interactions of 800 GeV/$c$ protons
with a beryllium oxide target. The resulting
particles pass through a series of collimators to produce
two nearly parallel beams. The beams also pass through lead and
beryllium absorbers to reduce the fraction of photons and neutrons
in each beam. Charged particles are removed from the beams by sweeping
magnets located downstream of the collimators.
To allow the $K_S$ component to decay away, the decay
volume begins approximately 94 meters downstream of the
target.  In this experiment we have two simultaneous
beams, one where the initial beam strikes
a regenerator composed of plastic scintillator and 
one that continues in vacuum.
For this analysis, we only consider
decays that originate in the beam opposite the
regenerator. The regenerator, located 125 meters
downstream of the target, is viewed by photomultiplier tubes which
are used to reject events that inelastically scatter
and deposit energy in the scintillator. 

The most critical detector elements for this analysis are a pure
CsI electromagnetic calorimeter\cite{csical} and a hermetic 
lead-scintillator photon veto system.
The CsI calorimeter is composed of 3100 blocks in a 1.9 m by 1.9 m array
that is 27 radiation lengths deep. Two 15 cm by 15 cm holes are located
near the center of the array for the passage of the two neutral beams.
For electrons with energies between 2 and 60 GeV,
the calorimeter energy resolution is below
1\% and the nonlinearity is less than 0.5\%. The position resolution
of the calorimeter is approximately 1 mm. The spectrometer is
surrounded by 10 detectors that veto photons at angles greater
than 100 milliradians. The first five vetos consist of 16
lead-scintillator layers with 0.5 radiation length lead sampling
followed by eight layers with 1.0 radiation length sampling.
The other five vetoes have 32 layers with 0.5 radiation length
sampling.
The most upstream photon veto is located 
just upstream of the regenerator. This module has two beam holes and
suppresses upstream decays. The remaining nine
vetoes are arrayed along the length of the detector
as shown in Figure~\ref{fig:det}. On the face of the CsI calorimeter
sits a tungsten-scintillator module that covers the
inner half of the CsI blocks surrounding the beam holes.  A
final photon veto consists of three
modules each 10 radiation lengths thick and sits behind the
CsI calorimeter, covering the beam holes. 
This photon veto is used to
reject events in which photons travel down 
either of the two beam holes in the calorimeter. 

The KTeV detector also contains a spectrometer for reconstructing
charged tracks. For this analysis, it is used to
calibrate the CsI calorimeter using electrons from 
$K_L\to\pi^\pm e^\mp\nu$ decays, and to veto events with charged particles.
This spectrometer consists of four planes of drift chambers; two located
upstream and two downstream of an analyzing magnet with
a transverse momentum kick of 0.4 GeV/$c$. 
Downstream of the CsI calorimeter, there is a 10 cm lead wall, followed by
a hodoscope (hadron-veto) used to reject hadrons hitting the calorimeter.

$K_L\to\pz\g\g$ events are recorded if they
satisfy certain trigger requirements. The event must
deposit greater than approximately 27 GeV in the CsI calorimeter and 
deposit no more than 0.5 GeV in the photon vetoes downstream of the
vacuum window located at 159 meters downstream of the target.
The trigger includes a hardware cluster processor\cite{hcc} 
that counts the number of contiguous clusters of blocks with
energies above 1 GeV. The total 
number of clusters is required to be exactly four. These trigger
requirements also select $K_L\to\pz\pz$ events that we use
as normalization for the $K_L\to\pz\g\g$ events.

Reconstructing $K_L\to\pz\g\g$ events is difficult since there
are few kinematical constraints giving rise to large backgrounds.
In particular, $K_L\to\pz\pz$ and $K_L\to\pz\pz\pz$ decays,
and hadronic interactions with material in the beam constitute 
the largest sources of background.
Backgrounds resulting from kaon decays with charged particles
are easily removed by discarding events with a large number of 
hits in the spectrometer. 

In the 
offline analysis, we require exactly four reconstructed clusters of
energy greater than 2.0 GeV while rejecting
events with additional clusters above 0.6 GeV.
Events with photons below 0.6 GeV do not contribute significantly
to the background. The total kaon energy is required to be
between 40 and 160 GeV.
Using the reconstructed energies and positions of the four clusters,
we calculate the kaon decay position, assuming that all four photons
result from the decay of a kaon. This position is then used to reconstruct a
$\pz$ from the six possible $\g\g$ pairings. We choose
the $\g\g$ pair whose reconstructed mass is closest to the nominal $\pz$ mass
and require that $|m_{12} - m_{\pz}|< 3.0 $ MeV/$c^2$. The 
$\pz$ mass resolution is 0.84 MeV/$c^2$. 

To remove $K_L\to\pz\pz$ decays we require
the mass $m_{34}$ be greater than 0.16 GeV/$c^2$ or less than 0.10 GeV/$c^2$.
We cut asymmetrically about the $\pz$ mass
to remove events with misreconstructed photons
near the calorimeter beam holes.
Since we use the best $\pz$ mass to choose among
the six possible $\g\g$ combinations, mispaired
$K_L\to\pz\pz$ events constitute a possible background. There
are three different ways to pair four photons to form
$\pz\pz$ combinations. We remove mispaired $K_L\to\pz\pz$
events by reconstructing the other two $\pz\pz$ pairings 
and discarding events that have two $\g\g$ combinations near
the $\pz$ mass.

The hadronic interaction background results from 
neutrons in the beam interacting with material, primarily the vacuum window
and the drift chambers.
Such interactions produce $\pz\pz$ and $\pz\eta$ pairs. These
events are not removed by the $2\pz$ cuts because the 
assumption that the invariant 
mass is $m_K$ is incorrect in this case.
To remove these events, we calculate the decay vertex of the
six $\g\g$ combinations assuming that two of the photons result
from a $\pz$ decay.
Using this decay vertex, we reconstruct the
mass of the other $\g\g$ pair and reject the event if the
decay vertex of this $\g\g$ pair reconstructs downstream of
158 meters and 
if the mass of that pair is within 15 MeV/$c^2$ of the known $\pz$ or 
$\eta$ mass. We also require little activity in the hadron-veto
to remove vacuum window interaction events that send hadrons into
the CsI calorimeter.

$K_L\to \pz\pz\pz$ decays are the most difficult source of background
to suppress in this analysis.
These events can 
contribute to the background primarily through the following
three mechanisms: a) four photons hit the calorimeter (zero fusion),
b) five photons hit the calorimeter with two overlapping or
fusing together to 
produce one cluster (single fusion), and c) all six photons hit the calorimeter and 
four photons fuse together to reconstruct as two photons (double fusion).
To reduce backgrounds from $K_L\to\pz\pz\pz$ events with 
zero and single fusions, we require little activity in the
photon veto detectors. 
These criteria are determined
from the amount of accidental activity in each counter and the
response of each counter to photons from $3\pz$ events.
For $K_L\to\pz\pz\pz$ events in which five or fewer photons
hit the calorimeter, the reconstructed decay vertex lies downstream
of the actual decay position. To remove these events, we require 
that the decay vertex be between 115 and 128 meters downstream of the
target, as shown in Figure~\ref{fig:det}.
To further reduce the number of events with missing
energy, we require that the center of energy of the
four photons lie within a 10 $\times$ 10 cm$^2$ region centered
on one of the beam holes in the CsI calorimeter.
The background from $3\pz$ events with only four photons
in the calorimeter is below 0.5\% of the expected signal level
after these cuts.

\begin{figure}[hbt]
  \centerline{
    \psfig{file=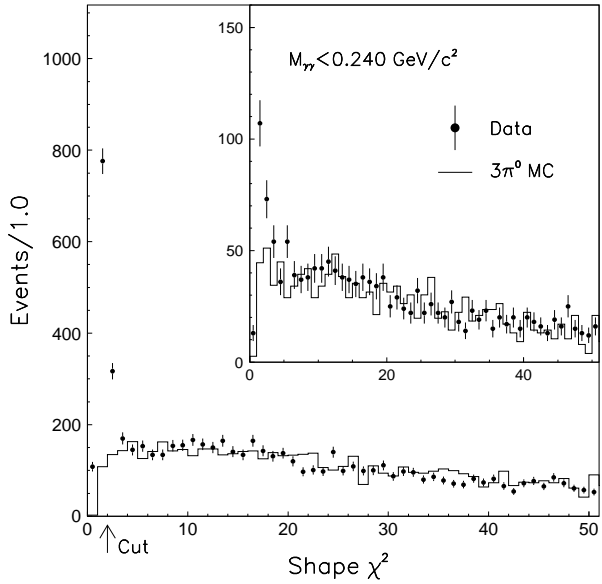,width=8.5cm}}
    \caption{The maximum shape $\chi^2$ after all requirements except
             for the shape $\chi^2$ cut. The dots are the data and the
             histogram is the $3\pz$ Monte Carlo. The excess at 
             low $\chi^2$ is due to $K_L\to\pz\g\g$ events and the 
             arrow indicates the position of our cut. The inset shows the
             same distribution for events with $m_{34}<0.240$ GeV/$c^2$,
             indicating the existence of a low-mass tail in $K_L\to\pz\g\g$
             events.}
  \label{fig:fuse3x3}
\end{figure}

The remaining $3\pz$ background results from
events in which two or more photons fuse together in the
CsI calorimeter. 
This background can be reduced by calculating a shape
$\chi^2$ for the energy deposited within the central
three-by-three array of CsI blocks of a cluster compared to
the energy distribution for a single photon.
Clusters from a single photon will have a low $\chi^2$,
whereas hadronic showers and fused clusters will usually result in
a large $\chi^2$. 
Figure~\ref{fig:fuse3x3} shows the distribution of the
maximum shape $\chi^2$ variable of the four photons
after all cuts have been imposed except for the shape $\chi^2$ requirement.
The $\pz\g\g$ signal reconstructs at low $\chi^2$, and the 
simulation of the $3\pz$ background agrees with the data above the
$\pz\g\g$ signal.
By requiring the maximum $\chi^2$ to be less than 2.0,
we are able to isolate a relatively clean sample of $K_L\to\pz\g\g$ events.
This requirement is effective in removing fused photons 
separated by more than 1 cm.

\begin{figure}[hbt]
  \centerline{
        \psfig{file=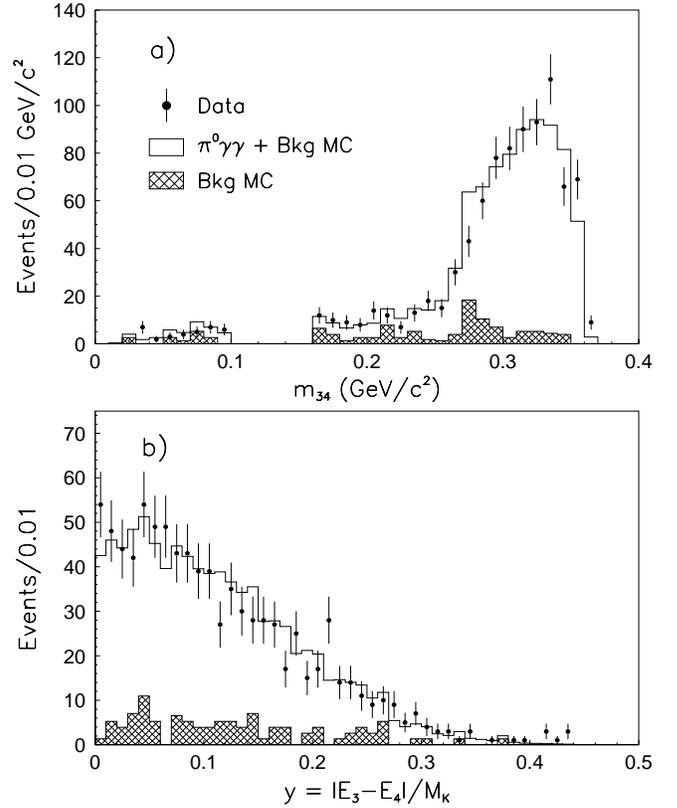,width=8.5cm}}
  \caption{The a) $m_{34}$ mass distribution and b) the $y$ parameter 
             for $K_L\to\pz\g\g$ candidates events. The
             dots are the data, the histogram is the sum of the background
             and $K_L\to\pz\g\g$ Monte Carlo and the dashed histogram is the
             normalized background Monte Carlo. The $K_L\to\pz\g\g$ MC is an
             ${\cal O}(p^6)$ calculation with $a_V = -0.7$.}
  \label{fig:candidates}
\end{figure}

Figure~\ref{fig:candidates}a shows the final $m_{34}$ distribution for all
events after making the photon shape $\chi^2$ requirement.
We find a total of 884 candidate events. Our simulation of the
background predicts $111\pm 12$ events dominated by 
$K_L\to\pz\pz\pz$ events as shown in Figure~\ref{fig:candidates}a.
The remaining $3\pz$ background is evenly divided between
double and single fusion events.
The level of the $3\pz$ background is determined by 
normalizing the $3\pz$ Monte Carlo to the shape $\chi^2$
distribution between 5 and 20 and is consistent with
absolutely normalizing the $3\pz$ events to the $2\pz$ events.
In this figure, we overlay the sum of the background
plus the $O(p^6)$ chiral perturbation prediction for $K_L\to\pz\g\g$. 
The shapes of the data and Monte Carlo calculation match very well.

Our event sample demonstrates for the first time the existence of
a low-mass tail in the $m_{34}$ distribution. The NA31 
experiment\cite{gbarr_pi0gg} had previously
set a limit of $\Gamma(m_{34}<0.240 \hbox{\rm\ GeV/}c^2)/
\Gamma(m_{34}\hbox{\rm\ all})<0.09$.  Our event sample
has $73\pm9\pm9$ events above a background of 
$47\pm 8\pm 5$ events in the region $m_{34}<0.240$ GeV/$c^2$.
This number corresponds to $\Gamma(m_{34}<0.240
\hbox{\rm\ GeV/}c^2)/\Gamma(m_{34}\hbox{\rm\ all})=12.7\pm1.3\pm1.5\%$
after correcting for events that are removed by the 
$K_L\to\pz\pz$ cut. Figure~\ref{fig:fuse3x3}
shows the maximum shape $\chi^2$ distribution 
for the events below 0.240 GeV/$c^2$
indicating an excess of events above the $3\pz$ background.

To extract a value for $a_V$, we perform a simultaneous fit
to the $m_{34}$ and $y$ distributions.
Figure~\ref{fig:candidates}b shows the $y$ distribution for our
final event sample.
From our fit, we obtain the value 
$a_V = -0.72\pm 0.05\pm0.06$.  
The systematic error is
dominated by our uncertainty in the $3\pz$ background.
We obtain a value of $a_V = -0.76\pm 0.09$ when we fit only the
$y$ distribution.
For the best fit to both distributions the
$\chi^2$ is 24.1 for 24 degrees of freedom.

To determine the $K_L\to\pz\g\g$ branching ratio, we normalize
the $K_L\to\pz\g\g$ events to $K_L\to\pz\pz$ events which helps
to reduce the systematic uncertainty in this measurement. The
acceptances for $\pz\g\g$ and $2\pz$ events are 
3.13\% and 3.23\%, respectively, for events with energies
between 40 and 160 GeV and decaying from 115 to 128 meters
downstream of the target. Applying the
same reconstruction criteria as those used in the $K_L\to\pz\g\g$
analysis but requiring $m_{34}$ between 130 MeV/$c^2$ and 140 MeV/$c^2$,
we find 441,309 $K_L\to\pz\pz$ events with negligible background.

Using a Monte Carlo in which $a_V$ is set to the best-fit
value, we extract the branching ratio for $K_L \to\pz\g\g$
by comparing the number of $\pz\g\g$ events to the number of
events in the normalization mode, $K_L \to\pz\pz$.  Systematic
uncertainties in this measurement come from 
the acceptance determination (2.4\%), 
the $2\pz$ branching ratio (2.1\%), 
the $3\pz$ background (2.1\%), 
the $a_V$ dependence (1.8\%),
the hadron veto requirement (1.8\%),
the photon shape requirement (1.8\%),
and the calibration (0.8\%).
The systematic uncertainties are added in quadrature, resulting in a
total systematic uncertainty of 5.0\%.
We find the branching ratio to be 
BR($K_L \to\pz\g\g$)
= $(1.68 \pm 0.07 \pm 0.08)\times 10^{-6}$.  
This is consistent with the ${\cal O}(p^6)$ prediction
\cite{dambrosio} 
with $a_V = -0.72 \pm 0.06$, which is $(1.53\pm0.10)\times 10^{-6}$.

In summary, our measurement of $K_L\to\pz\g\g$ decays shows the
first evidence of a low-mass tail in the $m_{34}$ distribution. 
This tail is predicted by
${\cal O}(p^6)$ chiral perturbation theory
calculations which include vector meson exchange. Our determination
of $a_V$ suggests that the CP-conserving contribution to $K_L \to\pz e^+e^-$
is between 1 and 2$\times 10^{-12}$
which is 2-3 orders of magnitude higher than predictions
based upon $O(p^4)$ calculations. The contribution of
the direct CP violating amplitude to the rate for $K_L\to\pz\ee$ 
is expected\cite{donoghue} to be between 1 and 4$\times 10^{-12}$.

We gratefully acknowledge the support and effort of the Fermilab
staff and the technical staffs of the participating institutions for
their vital contributions.  We also acknowledge 
G.~D'Ambrosio and F.~Gabbiani for useful discussions.
This work was supported in part by the U.S. 
Department of Energy, The National Science Foundation and The Ministry of
Education and Science of Japan.  In addition, A.R.B., E.B. and S.V.S. 
acknowledge support from the NYI program of the NSF; A.R.B. and E.B. from 
the Alfred P. Sloan Foundation; E.B. from the OJI program of the DOE; 
K.H., T.N. and M.S. from the Japan Society for the Promotion of
Science.  P.S.S. acknowledges receipt of a Grainger Fellowship.

%
%

\end{document}